\documentclass[journal]{IEEEtran}
\usepackage[normalem]{ulem}
\useunder{\uline}{\ul}{}
\usepackage{longtable}
\usepackage{amsmath}
\usepackage{amsfonts}
\usepackage{array}
\usepackage{bm}
\usepackage{framed}
\usepackage{tikz}
\usepackage{algorithmic}
\usepackage[algo2e]{algorithm2e} 
\usepackage{algorithm}
\usepackage{graphicx}
\usepackage{tabularx}
\usepackage{epstopdf}
\usepackage{amsthm}
\usepackage{mathabx}
\usepackage{amssymb}
\usepackage{subcaption}
\usepackage{color} 
\usepackage[numbers,sort&compress]{natbib}
\usepackage[margin=2cm]{geometry}
\usepackage{multirow}
\usepackage{alltt}
\usepackage[normalem]{ulem}
\usepackage[utf8]{inputenc}

\usepackage{pgfplots} 
\pgfplotsset{compat=newest} 
\pgfplotsset{plot coordinates/math parser=false} 
\newlength\figureheight 
\newlength\figurewidth 

\usepackage{bigstrut}
\usepackage{booktabs}
\usepackage{alltt}
\usepackage{multirow,booktabs}
\usepackage{multirow}

\theoremstyle{remark}

\definecolor{mycolor1}{rgb}{0.97255,0.97255,0.97255}%
\newcommand{\distas}[1]{\mathbin{\overset{#1}{\kern\z@\sim}}}%

\def \B {{\mathbf B}}

\def \Wt {\widetilde{W}}
\def \K {{\mathcal{K}}}
\def \N {{\mathcal{N}}}
\def \T {{\mathcal{T}}}
\def \S {{\mathcal{S}}}
\def \B {{\mathcal{B}}}
\def \L {{\mathcal{L}}}

\begin{document}
	
	%
	%
	\title{\vspace{7mm} Decentralized Multi-Antenna Coded Caching with Cyclic Exchanges}
	
	\author{Srujan Teja Thomdapu and Ketan Rajawat}
	
	\maketitle
	
	\begin{abstract}
	This paper considers a single cell multi-antenna base station delivering content to multiple cache enabled single-antenna users. Coding strategies are developed that allow for decentralized placement in the wireless setting. Three different cases namely, max-min multicasting, linear combinations in the complex field, and linear combinations in the finite field, are considered and closed-form rate expressions are provided that hold with high probability. For the case of max-min fair multicasting delivery, we propose a new coding scheme that is capable of working with only two-user broadcasts. A cyclic-exchange protocol for efficient content delivery is proposed and shown to perform almost as well as the original multi-user broadcast scheme.    
	\end{abstract}
	
	\section{Introduction}
	A tremendous growth in consumption of video over internet is driven the development of effective content delivery mechanisms seeking to maximize the user's quality of service. The wide-spread and heterogeneous nature of the traffic is the context of cellular systems exacerbates the content delivery problem. Caching popular content at the wireless edge is considered to be the most promising solution capable of enhancing the spectral efficiency \cite{liu2016caching} and reducing the load on the base station (BS). Generic content caching techniques are discussed in \cite{higgins2012informed} while further load reduction is possible if the cellular devices can communicate directly with each other \cite{hakola2010device}.  

Coded caching is a recent idea that allows file exchanges among the users thus reducing load on the server \cite{maddah2014fundamental}. Coded caching can be viewed as a virtual device-to-device communication network activated by carefully pre-placing content at the users. Coded caching algorithms for systems with multiple servers have been developed in \cite{shariatpanahi2016multi}. The centralized content placement routines are not entirely practical in wireless systems where the number of participating users fluctuates. A decentralized placement algorithm for wired systems is proposed in \cite{maddah2015decentralized} and has been further extended to hierarchical caching \cite{karamchandani2016hierarchical}, non-uniform demands \cite{niesen2017coded}, and online coded caching \cite{pedarsani2016online}.
	
Much less work has been done in the context of coded caching for wireless systems. The seminal work in \cite{shariatpanahi2017multi} put forth multi-antenna coded caching and has been extended to opportunistic content delivery \cite{ghorbel2017opportunistic}. Other related work can be found in \cite{girgis2017decentralized}. The existing schemes cannot be directly applied to practical cellular systems where the set of users available during the content delivery phase may be different from those accessible during the placement phase. Another practical issue surrounding multi-user broadcasts in cellular settings is that of user heterogeneity and fading. For instance, the broadcast rate decreases as the number of participating users increases. The issue is exacerbated in the next-generation millimeter wave systems that are highly directional \cite{niu2015survey}. As a simple demonstration, consider a set of 50 users randomly placed in a circle of unit radius according to a Poisson point process. Table \ref{pathloss} shows the average rate achieved when considering wireless broadasts from the BS located at the center of the disk. As can be seen from Table \ref{pathloss}, the broadcast rate for 3 users is almost 40\% below that obtained with two users. The reduction is even more drastic when 4 or more users is considered. Consequently, direct application of classical coded caching schemes suffers from the phenomenon of 'diminishing returns.' 

This paper puts forth a practical decentralized coded caching scheme that can work with only two-user broadcasts with negligible loss in performance. In particular, the first contribution of the paper is the development of the decentralized coded caching scheme for wireless systems. Building upon the results from \cite{shariatpanahi2017multi} for the centralized case, we derive closed-form expressions for the decentralized case. As in \cite{shariatpanahi2017multi}, we extend the delivery strategies by forming coded symbols in both complex and finite fields combined with Zero-Forcing. The second contribution is the new coding scheme that allows cyclic exchanges between users without requiring broadcasts among 3 or more users. The proposed scheme is shown to exploit more coding opportunities than a simple scheme with only two-user exchanges. 
		
	 Rest of the paper is organized as follows. Sec. \ref{sysmodel} briefs the system model. All the delivery strategies are presented in Sec. \ref{CDP}. Subsequently, the new proposed coding scheme is been presented in Sec. \ref{gcc}. Later we compare all the delivery schemes for decentralized placement and different coding schemes with max-min fair multicasting in Sec. 
	\ref{results}. Finally we conclude our paper in Sec. \ref{conclusion}. 
	
	\begin{table}[]
		\centering
		\begin{tabular}{|c|c|c|}
			\hline
			$R_2$ & $R_3$ & $R_4$ \\ \hline
			\textbf{0.2475}                                                        & \textbf{0.1425}                                               & \textbf{0.1004}                                                   \\ \hline
		\end{tabular}
		\caption{$K=50$, $L=1$ , $g = k_0\left(\frac{d}{d_0}\right)^{-3}$, $k_0=d_0=1$ and, $R_s = \mathbb{E}\left[ \min_{i \in S: S \subset \K,|S|=s}\log(1+g|\textbf{h}|^2)\right]$}
		\label{pathloss}
	\end{table}      
	
	\subsubsection*{\textbf{notations}} All capital bold letters represent matrices and all small bold letters represent vectors. $(.)^H$ denotes hermitian of the matrix/vector which is usually complex. $\mathbb{C}$ denote set of complex numbers, and $\oplus$ denotes bitwise xor operation. For any vector $\textbf{h}$, $\textbf{h}^{\perp}$ is a vector that present in the subspace which is perpendicular to $span(\textbf{h})$.           
	               
\section{System Model and Problem Formulation} \label{sysmodel}
This section details the system model under consideration and provides some background on the existing content delivery and caching approaches. Challenges arising in practical settings and the resulting problem formulation are also provided.

	\subsection{System Model}  
We consider a downlink scenario where a multi-antenna BS serves a set of cache-enabled mobile users $\mathcal{K}$ with $|\mathcal{K}| = K$. Each users is equipped with a single antenna while the BS is equipped with $L$ antennas. The BS has access to the server which contains the video library of $N$ files $\{W_1,...,W_N\}$ each of size $F$ bits. Each user has a fixed cache size that the BS can access during off-peak hours. The $F$-bit files are expressed as $f := F/m$ symbols in GF$(2^m)$. Each $m$-bit symbol is transmitted to the user over $n$ channel uses. To this end, the BS uses an encoder function $\psi:\mathbb{F}_{2^m}\rightarrow\mathbb{C}^n$, which encodes an $m$-bit symbol to $n$ complex numbers. The channel remains constant over the duration of transmission of these $n$ complex numbers. For the sake of brevity, we denote the encoded version of file $W$ as $\psi(W)=\Wt$ where the operator $\psi$ encodes a vector (row) of symbols element-wise. 

The concatenated wireless channel matrix between the BS and the users is represented by the matrix $\mathbf{H} \in \mathbb{C}^{L\times K}$. The BS has knowledge of the full CSIT. The received signal block over $n$ channel uses at user $k$ is given by
		\begin{align}
			\textbf{y}_k^H=\textbf{h}_k^H\textbf{X}+\textbf{z}_k^H
		\end{align}          		
$\textbf{y}_k\in\mathbb{C}^{n\times1}$ is the received signal block, $\textbf{z}_k\sim \mathcal{C}\mathcal{N}(0,I_n)$ is the additive white Gaussian noise, $\textbf{h}_k$ is the $k$-th column of $\mathbf{H}$, and $\textbf{X}\in\mathbb{C}^{L\times n}$ is the coded signal transmitted from the $L$ antennas over $n$ channel uses. The per-antenna transmit power constraint is given by 
		\begin{align}
			\frac{1}{nL}\sum_{i,j}\mathbb{E}\left[|\textbf{X}_{i,j}|^2\right]\leq P_{\max}.
		\end{align}
Without loss of generality, the received noise power is taken as unity. 

The content delivery takes place in two phases, namely placement phase and delivery phase. The placement phase takes place in advance and at off-peak hours. During the placement phase, the BS fills each user's cache with a fraction of the content library. However, the placed content may not be the same as the user's future demands, which are unknown at this stage. Delivery phase takes place at the peak hours where the BS employs coded caching to deliver the files to the users. The overall spectral efficiency depends on the user's caching capacity, henceforth denotes by $M$ files per user.

\subsection{Centralized Placement and Delivery}
We begin with discussing the centralized placement approach, first proposed in \cite{maddah2014fundamental} and utilized for wireless systems in \cite{shariatpanahi2017multi}. The technique in \cite{maddah2014fundamental} requires the total cached capacity per bit of content $t = MK/N$ to be an integer. Specifically, each file is partitioned into ${K \choose t}$ equal non-overlapping subfiles $W_{n,\T}$, where 
\begin{align}
\T &\subset \K  & |\T| = t
\end{align}
for all $n\in\N$. Equivalently, we have that 
\begin{align}
\bigcup_{\T\subseteq\K, |\T| = t} W_{n,\T} = W_n
\end{align}
Each subfile $W_{n,\T}$ is stored in all the caches of users in $\T$. The cache memory $MF$ is completely utilized with such placement, since $N {K-1 \choose t-1} = M {K\choose t}$. Further, each bit is shared by $t$ users. 

Let the user demands be collected into the vector $\textbf{d} = [d_1,...,d_K]^T$ where $d_k$ is the demand of the $k$-th user. The delivery proceeds as follows for each subset $\S \in \K$ where $|\S| = t+1$. The BS forms a coded message as, 
\begin{align} \label{centDel}
U_\S = \oplus_{k\in \S}W_{d_k, \S\setminus\{k\}}.
\end{align}
In other words, the constituent files are contained in all but one of the users in $\S$. In wired settings, all the coded messages are broadcast to all the users in $\K$ and each user decodes the relevant messages. 

In wireless settings, broadcast to a large number of users is wasteful. Instead, each coded message $U_\S$ is broadcast only to the users in $\S$. By utilizing Zero-Forcing and coding at different antennas, it is possible to exploit further multicasting opportunities as proposed in \cite{shariatpanahi2017multi}. 

Such a centralized placement scheme is however ill-suited to wireless settings where user mobility prevents the BS from carefully placing the subfiles to the set of users. For instance, if the number of active users change between the placement and delivery phases, using the coded messages in \eqref{centDel} would be highly suboptimal. Mobility is an important restriction in wireless settings and cannot be ignored. Such a restriction motivates the use of decentralized placement method, first proposed for wired systems in \cite{maddah2015decentralized}. Within the decentralized placement phase, each user is stuffed with randomly selected $MF/N$ bits of each file. While the decentralized placement scheme does not achieve the same delivery efficiency, the presence of each file at every user makes it robust to user mobility. Indeed, the content delivery phases does not depend on the number of users $K$, making the decentralized placement well-suited to wireless systems. 

The decentralized approach in \cite{maddah2015decentralized} is not applicable to the wireless system at hand. The next section details a delivery strategy that can be used in conjunction with the decentralized placement approach.

\section{Decentralized Placement and Delivery} \label{CDP}
Let $U_\S$ be the coded message transmitted by the BS to a subset $\S\subset \K$ of users. The length of $U_\S$ is denoted by $L\left(U_\S\right)F$ bits. Let $C\left(P_{\max},\S,\mathbf{H}\right)$bits/s denote the multi-cast rate at which BS transmits the common message to the users in subset $\S$. Hence the total transmission time to deliver all multi-cast sub-codewords is given by
\begin{align*}
T = \sum_{\S\subseteq\K}\frac{L\left(U_\S\right)F}{C\left(P_{\max},\S,\mathbf{H}\right)}.
\end{align*}
Here, the time $T$ is necessary for each user to decode the entire file $F$ bits. Likewise, the system's symmetric rate (good put) \cite{shariatpanahi2017multi} is given by 
\begin{align}\label{symRate}
R_{sym}=\frac{F}{T}=\left[ \sum_{\S\subseteq\K}\frac{L\left(U_\S\right)}{C\left(P_{\max},\S,\mathbf{H}\right)}\right]^{-1}.
\end{align}
The subsequent sections will provide detailed expressions for $R_{sym}$ for the proposed delivery strategies. In particular, we investigate three delivery strategies namely, max-min fair multicasting, linear combinations in complex field, and linear combinations in finite field. All these strategies have been investigated in \cite{shariatpanahi2017multi} for the centralized setting. In the present case however, we provide closed-forms for \eqref{symRate} for the decentralized case. 

\subsection{Max-Min fair multicasting} \label{mmd}
After the placement, denote every file $W$ as a collection of subfiles $\left(W_\S:\S\subseteq\K\right)$, where all the users in $\S$ cache the subfile $W_\S$. Let $V_{k,\S}$ denote the bits of the file $d_k$ cached exclusively at users in $\S$. That is, a bit of file $d_k$ is in $V_{k,\S}$ if it is present in the cache of every user in $\S$ and it is absent from the cache of every user outside $\S$. For multicast to the set $\S$ of users, the following sub-codeword is used:
\begin{align}\label{scw}
U_\S= \oplus_{k\in S}V_{k,\S\setminus\{k\}}.
\end{align} 
It is remarked that \eqref{scw} is similar to that used in the wired case \cite{maddah2015decentralized} though its impact would be different in the wireless setting. Different from the centralized placement, the randomized placement only allows us to obtain an expression for $R_{sym}$ that holds with high probability. Indeed, due to the random placement, the sub-codeword length ($L\left(U_\S\right)F$ bits) adheres to the following approximation:
		\begin{align}\label{scwl}
			L\left(U_\S\right) \approxeq \left(\frac{M}{N}\right)^{|\S|-1} \left(1-\frac{M}{N}\right)^{N-|\S|+1}.	
		\end{align}
	which holds with high probability \cite{maddah2015decentralized}. Concatenating these sub-codewords yields the coded multicast codeword given by,
		\begin{align} \label{concCwd}
		X = \{U_\S:|\S|=s\}_{s=1}^K
		\end{align}  
It is remarked that \eqref{concCwd} is different from that obtained for the centralized case\cite{shariatpanahi2017multi}. 

The message $U_\S$ is useful only to the users in $\S$. Hence BS transmits $U_\S$ by multicasting to the users in $\S$. Since the CSIT is available, BS uses a beamforming vector $\textbf{w}_\S$ and transmits with common rate as,
		\begin{align} \label{maxminS}
		R(\S)=\min_{k\in \S}\log\left(1+|\textbf{h}_k^H\textbf{w}_\S|^2 P_{\max}\right)
		\end{align}
which can be maximized by choosing,
		\begin{align}
		\textbf{w}_\S^{\star} = &\arg\max_{\textbf{w}}\min_{k\in \S}|\textbf{h}_k^H\textbf{w}|, \label{maxmin}\\
		\text{s.t}.\hspace{3mm} &\left \|\textbf{W}\right \|\leq 1.
		\end{align}
		Now the symmetric rate expression for this strategy can be written as 
		\begin{align*}
			R_{sym} = \left(\sum_{\{\S:\S\in\K,|\S|\geq 1\}}\frac{L(U_\S)}{R(\S)}\right)^{-1}.
		\end{align*} 
It is remarked that the max-min fair multicasting case is quite similar to the wired case with decentralized placement. However, the same does not hold for the subsequent two subsections.

		 \begin{algorithm}
			\caption{Delivery phase in complex field}
			\begin{algorithmic}
				\STATE Demands (\textbf{d}) are given. 
				\STATE \For{$s = 1,2,..K$}
				{\STATE INDEX-INIT
				{\STATE \For{$\B\subseteq\K$, $|\B| = \min\{s+L-1,K\}$}
						{\STATE\For{$\S\subseteq\B$, $|\S| = s$}
							{\STATE $\textbf{u}_{\B}^{\S} = BFV(\B,\S,\mathbf{H})$}
						\STATE $\boldsymbol{\omega} = {\min\{s+L-1,K\}-1 \choose s-1}$
						\STATE \For{$\omega = 1,2,...,\boldsymbol{\omega}$}
							{\STATE \For{$\S\subseteq\B$, $|\S| = s$}
								{$G_{\omega}(\S) \leftarrow \L^{\omega}_{k\in\S}\left(\Wt_{d_k, \S\setminus\{k\}}^{N(k,\S\setminus\{k\})}\right)$.}
							\STATE $\textbf{X}_{\omega}(\B)\leftarrow \sum_{\S\subseteq\B,|\S|=s}\textbf{u}_{\B}^{\S}G_{\omega}(\S)$}
							\STATE  Transmit $\textbf{X}(\B) = \left[\textbf{X}_1(\B), ..., \textbf{X}_{\boldsymbol{\omega}}(\B)\right]$.
							\STATE INDEX-UPDATE}}}			
			\end{algorithmic}
			\label{lccf}
		\end{algorithm}
	
		\begin{algorithm}
			\caption{Auxiliary procedures}
			\begin{algorithmic}			
				\STATE \textbf{procedure} $\textbf{u}_{\B}^{\S} = BFV(\B,\S,\mathbf{H})$
				\STATE \hspace{2mm} \textbf{if} $|\S|\leq K-L+1$
				\STATE \hspace{4mm} Select $\textbf{u}_{\B}^{\S}$ such that,
				\STATE \hspace{6mm}$\textbf{u}_{\B}^{\S} \perp \textbf{h}_k\text{ }\forall k\in \B\setminus \S$ 
				 \STATE \hspace{6mm} $\textbf{u}_{\B}^{\S} \not\perp \textbf{h}_k\text{ }\forall k\in \S$
				\STATE \hspace{2mm} \textbf{else}
				\STATE \hspace{4mm} $\textbf{u}_{\B}^{\S} = \arg\max_{\textbf{u}} \min_{k\in \S}\textbf{h}_k^H\textbf{u} \hspace{3mm} \text{s.t } \textbf{h}_k^H\textbf{u}=0 \forall k \in \B\setminus \S$
				\STATE \hspace{2mm} \textbf{end}
				\STATE \textbf{end procedure}
				
				\STATE \textbf{procedure INDEX-INIT}
				\STATE \hspace{2mm} \textbf{for} $\S\subseteq\K$, $|\S| = s$
				\STATE \hspace{4mm} $N(k,\S\setminus\{k\}) \leftarrow 1$ $\forall k \in \S$
				\STATE \hspace{2mm} \textbf{end}
				\STATE \textbf{end procedure}
				
				\STATE \textbf{procedure INDEX-UPDATE}
				\STATE \hspace{2mm} \textbf{for} $\S\subseteq\B$, $|\S| = s$
				\STATE \hspace{4mm} $N(k,\S\setminus\{k\}) \leftarrow N(k,\S\setminus\{k\}) + 1$ $\forall k \in \S$
				\STATE \hspace{2mm} \textbf{end}
				\STATE \textbf{end procedure}
 			\end{algorithmic}
			\label{auxpro}
		\end{algorithm}
		
\subsection{Linear combination in the complex field}		
The symbols are first converted into complex numbers and then coded by forming linear combinations among different sub-files before transmission to the users. We consider a simple example that helps elucidating this aspect. 
		\subsubsection*{Example A}
		Let us consider a BS with $L=2$ transmit antennas, $K=3$ users, $N=3$ files $A,B,C$, encoded versions are $\tilde{A},\tilde{B},\tilde{C}$. Each user collects $(M/3)f$ symbols from each file. Now we divide the each file $\Wt\in\{\tilde{A},\tilde{B},\tilde{C}\}$ into subfiles based on the placement as below.
		\begin{align*}
		\Wt\rightarrow\left(\Wt^1_0,\Wt^2_0,\Wt_1,\Wt_2,\Wt_3,\Wt_{12},\Wt_{13},\Wt_{23},\Wt_{123}\right)
		\end{align*}
		where $\Wt_{\S}$ is the subfile that's been stored elusively in the caches of users $\{k : k\in\S\}$ and, the subfile $\Wt_0$ is further divided into two equal subfiles that represented as $\Wt_0^1,\Wt_0^2$. Note that, these subfile division is entirely different from \cite{shariatpanahi2017multi} in which the size of each subfile is equal and it is stored exclusively in corresponding subsets of users. In contrast, the sub-files here are of different sizes and stored at the users of all possible subsets of users. Existence of all possible subset of users is an outcome of the decentralized placement. Another key difference here is division of subfiles takes place before/after the caching in centralized/decentralized placement. 
		
		Now let us say each user requests $A,B,C$ respectively.
		\subsubsection*{s=1}
		Now consider $2$-subset of users $\{1,2\}$, $\{2,3\}$, $\{1,3\}$ and, the message blocks are,
		\begin{align*}
		\textbf{X}(\{1,2\}) =& \frac{1}{\sqrt{2}}\left[\tilde{A}^1_0\frac{\textbf{h}_2^{\perp}}{\left \|\textbf{h}_2^{\perp}\right\|}+\tilde{B}^1_0\frac{\textbf{h}_1^{\perp}}{\left \|\textbf{h}_1^{\perp}\right\|}\right] \\
		\textbf{X}(\{2,3\})  =&\frac{1}{\sqrt{2}}\left[\tilde{B}^2_0\frac{\textbf{h}_3^{\perp}}{\left \|\textbf{h}_3^{\perp}\right\|}+\tilde{C}^1_0\frac{\textbf{h}_2^{\perp}}{\left \|\textbf{h}_2^{\perp}\right\|}\right] \\
		\textbf{X}(\{1,3\}) =&\frac{1}{\sqrt{2}}\left[\tilde{A}^2_0\frac{\textbf{h}_3^{\perp}}{\left \|\textbf{h}_3^{\perp}\right\|}+\tilde{C}^2_0\frac{\textbf{h}_1^{\perp}}{\left \|\textbf{h}_1^{\perp}\right\|}\right]
		\end{align*}
		We present decoding of the message block in favor of user $1$, and for other users the procedure is similar. Decoding $(\tilde{A}_0^1,\tilde{A}_0^2)$ is straight forward. The transmission rates $R(\{1,2\},1)$ for $\tilde{A}_0^1$, $\tilde{A}_0^2$ are, $\log\left(1+\frac{P_{\max}}{2}\min\left(\frac{|\textbf{h}_1^H\textbf{h}_2^{\perp}|^2}{\left \|\textbf{h}_2^{\perp}\right\|^2},\frac{|\textbf{h}_2^H\textbf{h}_1^{\perp}|^2}{\left \|\textbf{h}_1^{\perp}\right\|^2}\right)\right)$, $\log\left(1+\frac{P_{\max}}{2}\min\left(\frac{|\textbf{h}_1^H\textbf{h}_3^{\perp}|^2}{\left \|\textbf{h}_3^{\perp}\right\|^2},\frac{|\textbf{h}_3^H\textbf{h}_1^{\perp}|^2}{\left \|\textbf{h}_1^{\perp}\right\|^2}\right)\right)$. Similarly one can find out for other sets.
		\subsubsection*{s=2}
		Now consider $3$-subset of users, which is only one $\B=\{1,2,3\}$. The message block with two sequential transmissions  is written as $[\textbf{X}_1(\B), \textbf{X}_2(\B)]$. Where $\textbf{X}_1(\B)$ is, 
		\begin{align*}
		 \frac{1}{\sqrt{6}}\left[(\tilde{A}_2+\tilde{B}_1)\frac{\textbf{h}_3^{\perp}}{\left \|\textbf{h}_3^{\perp}\right\|}+(\tilde{B}_3+\tilde{C}_2)\frac{\textbf{h}_1^{\perp}}{\left \|\textbf{h}_1^{\perp}\right\|}+(\tilde{A}_3+\tilde{C}_1)\frac{\textbf{h}_2^{\perp}}{\left \|\textbf{h}_2^{\perp}\right\|}\right]
		\end{align*}
		and, $\textbf{X}_2(\B)$ is, 
		\begin{align*}
		 \frac{1}{\sqrt{6}}\left[(\tilde{A}_2+\tilde{B}_1)\frac{\textbf{h}_3^{\perp}}{\left \|\textbf{h}_3^{\perp}\right\|}+(\tilde{C}_2-\tilde{B}_3)\frac{\textbf{h}_1^{\perp}}{\left \|\textbf{h}_1^{\perp}\right\|}-(\tilde{A}_3+\tilde{C}_1)\frac{\textbf{h}_2^{\perp}}{\left \|\textbf{h}_2^{\perp}\right\|}\right]
		\end{align*}
		With help of its cache contents, the first user extracts
		\begin{align*}
		\frac{1}{\sqrt{3}}U\begin{bmatrix} \frac{\textbf{h}_1^H\textbf{h}_3^{\perp}}{\left \|\textbf{h}_3^{\perp}\right\|} &0 \\ 0 & \frac{\textbf{h}_1^H\textbf{h}_2^{\perp}}{\left \|\textbf{h}_2^{\perp}\right\|}\end{bmatrix}\begin{bmatrix} \tilde{A}_2 \\\tilde{A}_3 \end{bmatrix}+\begin{bmatrix} \textbf{z}_1^1\\\textbf{z}_1^2 \end{bmatrix}
		\end{align*}
		Where $U$ is the unitary matrix represented as, $1/\sqrt{2}\begin{bmatrix}1&1 \\1 &-1 \end{bmatrix}$. User $1$ decodes with a rate less than $R(\{1,2,3\},1,2)=\log\left(1+\frac{P_{\max}}{3}\min\left(\frac{|\textbf{h}_1^H\textbf{h}_2^{\perp}|^2}{\left \|\textbf{h}_2^{\perp}\right\|^2},\frac{|\textbf{h}_1^H\textbf{h}_3^{\perp}|^2}{\left \|\textbf{h}_3^{\perp}\right\|^2}\right)\right)$. And the common rate for all the users in $\{1,2,3\}$ is,
		\begin{align*}
		R(\{1,2,3\},2)=\log\left(1+\frac{P_{\max}}{3}\min_{i,j\in\{1,2,3\},i\neq j}\left(\frac{|\textbf{h}_i^H\textbf{h}_j^{\perp}|^2}{\left \|\textbf{h}_j^{\perp}\right\|^2}\right)\right)
		\end{align*}
		\subsubsection*{s=3}
		Now consider the whole set $\{1,2,3\}$. Now the message block is,
		\begin{align*}
		\textbf{X}(\{1,2,3\}) = \frac{1}{\sqrt{3}}\left[\left(\tilde{A}_{23}+\tilde{B}_{13}+\tilde{C}_{12}\right)\frac{\textbf{u}_{\{1,2,3\}}^{\{1,2,3\}}}{\left \|\textbf{u}_{\{1,2,3\}}^{\{1,2,3\}}\right\|}\right]
		\end{align*}
		Where $\textbf{u}_{\{1,2,3\}}^{\{1,2,3\}}$ is the solution of \eqref{maxmin}. Here the common rate for all the users is,
		\begin{align*}
		R(\{1,2,3\},3)=\log\left(1+\frac{P_{\max}}{3}\min_{i\in\{1,2,3\}}\left(\frac{\left|\textbf{h}_i^H\textbf{u}_{\{1,2,3\}}^{\{1,2,3\}}\right|^2}{\left \|\textbf{u}_{\{1,2,3\}}^{\{1,2,3\}}\right\|^2}\right)\right)
		\end{align*}
		The example can be generalized as shown in Algorithm \ref{lccf} which is explained in Appendix \ref{gen}. Algorithm \ref{lccf} is inspired from Algorithm 1 in \cite{shariatpanahi2017multi} and provides the delivery strategy for the decentralized placement. The expression for symmetrical rate is provided in Appendix \ref{srlccf}
		
		\subsection{Linear combination in the finite field}
		\begin{algorithm}
			\caption{Delivery phase in finite field}
			\begin{algorithmic}
				
				\STATE Demands (\textbf{d}) are given. 
				\STATE INDEX-INIT
				\STATE \For{$s = 1,2,..K$}
				{\STATE \For{$\B\subseteq\K$, $|\B| = \min\{s+L-1,K\}$}
					{\STATE\For{$\S\subseteq\B$, $|\S| = s$}
						{\STATE $\textbf{u}_{\B}^{\S} = BFV(\B,\S,\mathbf{H})$
						\STATE $G_{\omega}(\S) \leftarrow \oplus_{k\in\S}\left(W_{d_k, \S\setminus\{k\}}^{N(d_k,\S\setminus\{k\})}\right)$.}
						\STATE $\boldsymbol{\omega} = {\min\{s+L-1,K\} \choose s}$
							\STATE $\textbf{X}(\B)\leftarrow \sum_{\S\subseteq\B,|\S|=s}\frac{\textbf{u}_{\B}^{\S}}{\sqrt{\boldsymbol{\omega}}}\psi \left(G_{\omega}(\S)\right)$
						\STATE  Transmit $\textbf{X}(\B) = \left[\textbf{X}_1(\B), ..., \textbf{X}_{\boldsymbol{\omega}}(\B)\right]$ 
						\STATE INDEX-UPDATE}}			
			\end{algorithmic}
			\label{lcff}
		\end{algorithm}
		The strategy exactly similar to the one in complex field except that the symbols are first coded and then converted into complex numbers. Let us examine this with our example.
		\subsubsection*{Example A} We explain for $s = 2$. For $s=1$, and $s=3$, the expressions are straight forward and exactly similar to the one in complex field. The message block for $s = 2$ is to the users $\B = \{1,2,3\}$ can be written as,
		\begin{align*}
			\textbf{X}(\B) = \frac{1}{\sqrt{3}}\Bigg[&\psi(A_2\oplus B_1)\frac{\textbf{h}_3^{\perp}}{\left \|\textbf{h}_3^{\perp}\right\|}+\psi(B_3\oplus C_2)\frac{\textbf{h}_1^{\perp}}{\left \|\textbf{h}_1^{\perp}\right\|}\\
			+&\psi(A_3\oplus C_1)\frac{\textbf{h}_2^{\perp}}{\left \|\textbf{h}_2^{\perp}\right\|}\Bigg].	
		\end{align*} 
		User $1$ receives the following message,
		\begin{align*}
			\psi(A_2\oplus B_1)\frac{\textbf{h}_1^H\textbf{h}_3^{\perp}}{\sqrt{3}\left \|\textbf{h}_3^{\perp}\right\|} + \psi(A_3\oplus C_1)\frac{\textbf{h}_1^H\textbf{h}_2^{\perp}}{\sqrt{3}\left \|\textbf{h}_2^{\perp}\right\|} +  \textbf{z}_1.
		\end{align*}
		User decodes both symbols by considering MAC channel and, the rates are defined as,
		\begin{align}
			\hspace{-3mm} R_{Sum}(\B,1,2) & = \log\left(1 + \frac{1}{3} \left( \frac{|\textbf{h}_1^H\textbf{h}_3^{\perp}|^2}{\left \|\textbf{h}_3^{\perp}\right\|^2}   + \frac{|\textbf{h}_1^H\textbf{h}_2^{\perp}|^2}{\left \|\textbf{h}_2^{\perp}\right\|^2} \right)P_{\max}\right) \label{sumrate}\\
			R_{\{1,2\}}(\B,1,2) & = \log\left(1 + \frac{1}{3} \left( \frac{|\textbf{h}_1^H\textbf{h}_3^{\perp}|^2}{\left \|\textbf{h}_3^{\perp}\right\|^2}  \right)P_{\max}\right) \label{1,2} \\ 
			R_{\{1,3\}}(\B,1,2) & = \log\left(1 + \frac{1}{3} \left( \frac{|\textbf{h}_1^H\textbf{h}_2^{\perp}|^2}{\left \|\textbf{h}_2^{\perp}\right\|^2} \right)P_{\max}\right) \label{1,3}
		\end{align}
The effective sum rate for user 1 is given by $R_{Eff}(\{1,2,3\},1,2) = \min(\text{Eq. }\eqref{sumrate}, 2 \times \text{Eq. }\eqref{1,2}, 2 \times \text{Eq. }\eqref{1,3})$. The  common rate for all the users in $\{1,2,3\}$ is minimum of all the  effective sum rates for all the users. The generalization is shown in Algorithm \ref{lcff} which is essentially an extension of \cite[Algorithm 2]{shariatpanahi2017multi} such that it applies for the decentralized placement setting. The expression for symmetrical rate for this strategy is derived in Appendix \ref{srlcff}.  
		
\section{Decentralized Coded Caching - Cyclic Exchanges with Two User Multicast} \label{gcc}
Achieving high broadcast rate to multiple wireless users is not possible due to the heterogeneity among users, both deterministic (path-loss) and random (small-scale fading). For instance, if one of the users is far  from side the BS, the broadcast rate to a set of users would always be low. This section provides with a novel technique that achieves a high spectral efficiency while requiring only two-user multicast messages. The idea is to emulate broadcasts using cyclic exchanges between multiple users. In order to demonstrate the idea, we consider it within the gamut of max-min fair multicasting. 

The placement phase remains the same as in the previous section. In the content delivery phase, once the user demands are revealed, we construct a directed graph $\mathcal{G}= (\mathcal{K},\mathcal{E})$, with set of nodes $\mathcal{K}$  and set of edges $\mathcal{E}$. Let $\mathcal{K}$ be the set of users that demanded files. The edge $e_{ij}$ is the number of symbols that are stored in the cache of user $i$ and not in cache of user $j$ that are demanded by user $j$. It is possible that the symbols demanded by user $j$ are stored in multiple caches of other users. Then to avoid redundancy any one of that users is randomly selected to make an edge. We remark that the graph construction process is not unique. The BS transmits one symbol at a time and the edges are updated for every symbol transmission and so is the graph. All loops $\L = \{(l,O_l)\}$ in the graph are first extracted and code words are generated for every loop. Each loop can be characterized as $(\mathcal{K}_l,\mathcal{E}_l)\subset \mathcal{G}$, where $\mathcal{K}_l$, $\mathcal{E}_l$ represent set of nodes and  edges that present in the loop $l$ respectively. The number of nodes in $\mathcal{K}_l$ is the order $O_l$ of the loop $l$. All the nodes in $\mathcal{K}_l$ are arranged in a way that the edge is directed from node in left side to node in right side  and, loop ends with an edge from last node to first one. 
		 
		 \begin{algorithm}
		 	\caption{Coded Caching - Cyclic exchanges with two user multicasts}
		 	\begin{algorithmic}
		 		\STATE Demands (\textbf{d}) are given. 
		 		\STATE Construct a directed graph $\mathcal{G} = (\mathcal{K},\mathcal{E})$ as explained in \ref{gcc}.
		 		\STATE Extract all the loops $\L = \{(l,O_l)\}$ in graph $\mathcal{G}$. 
		 		\STATE \For{o = 2,3,...,K} 
		 		{\For{$\{(l,O_l)\} \subset \L$ : $O_l = o$}
		 			{\If{loop $l = (\mathcal{K}_l,\mathcal{E}_l)$ exists}{ \STATE select any user $u\in\mathcal{K}_l$ and the edge $e_{uv}\in\mathcal{E}_l$ that strats from $u$. 
		 					\STATE $k=v$
		 					\STATE \While{$k\neq u$}
		 					{\STATE Select an edge $e_{kj}$ such that it starts from $k$.
		 						\STATE Transmit $U_{\{u,k\}} = \Wt_{d_k}\oplus\Wt_{d_j}$ with the rate in \eqref{maxminS}. 
		 						\STATE $k = j$}
		 					\STATE \textbf{Update} $e_{ij} \leftarrow e_{ij} -1$ $\forall e_{ij}\in\mathcal{E}_l$. }}}
		 		\STATE Remaining symbols are uncoded transmissions.
		 	\end{algorithmic}
		 	\label{maxmingcc}
		 \end{algorithm}
	 
	 	 \begin{figure}
	 		\setcounter{subfigure}{0}
	 		\begin{subfigure}{\columnwidth}
	 			\includegraphics[width=1.005\linewidth, height = 0.65\linewidth]
	 			{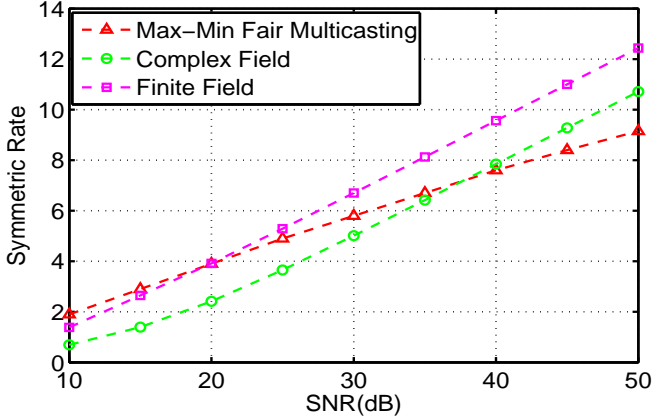}
	 			\caption{Centralized Placement Scheme}
	 			\label{cen}
	 		\end{subfigure}
	 		\begin{subfigure}{\columnwidth}
	 			\vspace{15mm}
	 			\includegraphics[width=1.005\linewidth,height = 0.65\linewidth]
	 			{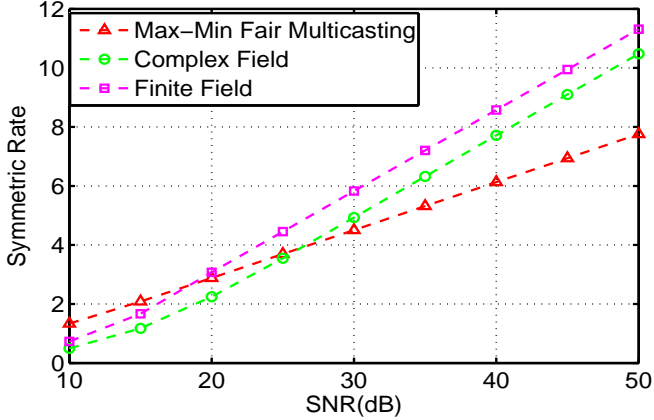}
	 			\caption{Decentralized Placement Scheme }
	 			\label{decen}
	 		\end{subfigure}
	 		\caption{Comparison results among different delivery schemes for the example $K=3$, $N=3$, $L=2$ and, $M=1$.}
	 		\label{DMACC}
	 	\end{figure} 
	 
		\subsection*{Example B : $\left(\{1,2,3\},\{e_{12},e_{23},e_{31}\}\right)$}  
		\begin{center}
			\begin{tikzpicture}
			\draw [thick] (-1,0) node[]{$\Wt_{d_2}$} circle(0.5)  (1,0) node[]{$\Wt_{d_3}$} circle(0.5)  (0,-2) node[]{$\Wt_{d_1}$} circle(0.5) (-1,0.75) node[]{1} (1,0.75) node[]{2} (0,-2.75) node[]{3};
			\draw [very thick,->] (-0.74,0) -- (0.74,0);
			\draw [very thick,->]  (1,-0.25) -- (0,-1.75);
			\draw [very thick,->]  (-0.1,-1.75) -- (-1,-0.25);
			\draw [thick] (0,0) node[above]{$e_{12}$} (0.5,-1) node[right]{$e_{23}$} (-0.5,-1) node[left]{$e_{31}$}
			;\end{tikzpicture}
		\end{center}  
		Note that, there are more than one way to construct code words. One possible way of construction is shown below.
		\begin{align}
			U_{\{1,3\}} &= \Wt_{d_1} \oplus \Wt_{d_2} \\ 
			U_{\{2,3\}} &= \Wt_{d_2} \oplus \Wt_{d_3}.
		\end{align}
		user 3 first retrieve $\Wt_{d_2}$ followed by its demand $\Wt_{d_3}$. All users maintain a temporary memory that can store one symbol. 
		
		\subsection*{Example C : $\left(\{1,2,3,4\},\{e_{12},e_{23},e_{34},e_{41}\}\right)$} 
		\begin{center}
			\begin{tikzpicture}
			\draw [thick] (-1,0) node[]{$\Wt_{d_2}$} circle(0.5)  (1,0) node[]{$\Wt_{d_3}$} circle(0.5) (-1,-2) node[]{$\Wt_{d_1}$} circle(0.5) (1,-2) node[]{$\Wt_{d_4}$} circle(0.5) (-1,0.75) node[]{1} (1,0.75) node[]{2} (-1,-2.75) node[]{4} (1,-2.75) node[]{3};
			\draw [very thick,->] (-0.74,0) -- (0.74,0);
			\draw [very thick,->] (0.74,-2) -- (-0.74,-2);
			\draw [very thick,->]  (1,-0.25) -- (1,-1.75);
			\draw [very thick,->]  (-1,-1.75) -- (-1,-0.25);
			\draw [thick] (0,0) node[above]{$e_{12}$} (0,-2) node[below]{$e_{34}$} (1,-1) node[right]{$e_{23}$} (-1,-1) node[left]{$e_{41}$}
			;\end{tikzpicture}
		\end{center}   
		The possible codewords are written as,
		\begin{align}
			U_{\{1,4\}} &= \Wt_{d_1} \oplus \Wt_{d_2} \\ 
			U_{\{2,4\}} &= \Wt_{d_2} \oplus \Wt_{d_3} \\
			U_{\{3,4\}} &= \Wt_{d_3} \oplus \Wt_{d_4}.	
		\end{align}
	 
	   All the first order loops are direct exchanges. For example, the loop $\left(\{1,2\},\{e_{12},e_{21}\}\right)$ has exactly one possible code word and that is, $U_{\{1,2\}} = \Wt_{d_1} \oplus \Wt_{d_2}$.  There are no transmissions for a loop with order 1 which is self contained. Symbols can be retrieved directly by node itself. The generalized algorithm is been presented in Algorithm \ref{maxmingcc}.
	       
	  \section{Results} \label{results}
  
	  Simulations are performed for Example A and compared all the schemes by changing SNR = $P_{\max}$ values. Results in Fig. \ref{decen} represent the different delivery schemes for decentralized placement and the results in Fig. \ref{cen} represent for centralized placement \cite{shariatpanahi2017multi}. The performance is good for centralized placement which is as expected. One can easily verify that the both analytical and numerical results are consistent with \cite{maddah2015decentralized} and \cite{shariatpanahi2017multi}.
	  
\begin{figure}
	\setcounter{subfigure}{0}
	\begin{subfigure}{\columnwidth}
		\includegraphics[width=1.005\linewidth, height = 0.65\linewidth]
		{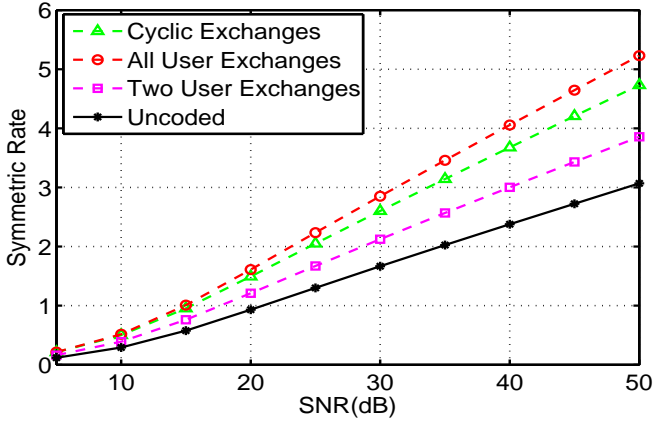}
		\caption{Homogeneous Placement with $M=1.5$}
		\label{gcc_homo}
	\end{subfigure}
	\begin{subfigure}{\columnwidth}
		\vspace{15mm}
		\includegraphics[width=1.005\linewidth,height = 0.65\linewidth]
		{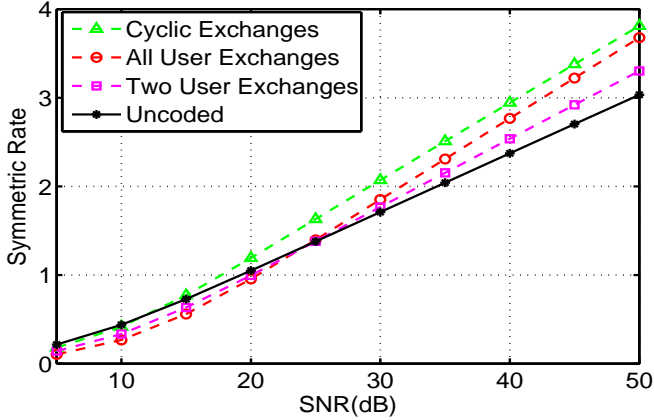}
		\caption{Heterogeneous Placement with $M=\{0.5,1,1.5,2,2.5\}$ }
		\label{gcc_hetero}
	\end{subfigure}
	\caption{Comparison results among different coding schemes with Max-Min Fair Multicasting delivery for the example $K=5$, $N=5$ and, $L=2$.}
	\label{DMACC_GCC}
\end{figure}

	  The comparisons for various coding schemes with max-min fair multicast delivery are showed in Fig.\ref{DMACC_GCC}. For this simulation, we considered a setting in which all the users are placed in an unit circle following Poisson point process. And the attenuation factor due to the path-loss is considered to be $k_0(d_k/d_0)^{-3}$, where $d_k$ is the distance between BS and the user $k$ and $k_0$, $d_0$ are the parameters which are considered to be 1. The 'Two User Exchanges' is the coding scheme similar to \cite{maddah2015decentralized} in which the symbols that are involved with multi ($\geq 2$) user exchanges are considered as uncoded transmissions. Heterogeneous caching is also considered for comparisons (see Fig. \ref{gcc_hetero}), in which the users place contents randomly and the implementation of coding scheme is exactly similar to the one in \cite{maddah2015decentralized}. The proposed scheme beats the two user exchanges scheme which enlightens the fact that more coding opportunities are can be exploited with cyclic exchanges in both cases. At low SNR values the proposed scheme performs as good as 'All User Exchanges'. For heterogeneous caching case (Fig. \ref{gcc_hetero}), the scheme with cyclic exchanges outperforms all kind of coding schemes. Interestingly uncoded scheme beats all schemes at low SNR which is due to the less common rate for multicasting delivery and lower coding opportunities. 
	 
	  \section{Conclusion} \label{conclusion}
 This paper considered a downlink transmission to cache enabled users via multiple antenna base station with decentralized placement scheme at user's terminals. We have extended all the analytical results in \cite{shariatpanahi2017multi} to new placement strategy and including to this we propose a new coding scheme which is called cyclic exchanges on top of max-min fair multicasting by restricting to only two user multicasting. From simulation results we showed the proposed scheme outperforms the 'Two User Exchanges' and performs as good as the 'All User Exchanges'.  
 
 \section*{Acknowledgment}
 The results were obtained at SPiN lab in Indian Institute of Technology Kanpur with the financial support of "Intel India PhD Fellowship Program" sponsored by Intel Technology India Pvt Ltd.         
	   
\appendices
\section{Generalization} \label{gen}
\subsection{Division of Subfiles:}
The cache content placement works exactly like in \cite{maddah2015decentralized}, but here each user stores a random floor(${Mf/N}$) symbols from each file of size $f$ symbols. Denote every file $\psi(W_n)=\Wt_n$ as a collection of subfiles $\left(\Wt_{n,\S}:\S\subseteq\K\right)$, obtained by encoding all the cached symbols with the encoding function $\psi(.)$. We divide the above subfiles into ${K-s-1\choose L-1}$ non-overlapping equal-sized mini-files as follows:
\begin{align*}
\Wt_{n,\S} = \left(\Wt_{n,\S}^j:j=1,...,{K-s-1\choose L-1}\right) \forall \S\subseteq\K
\end{align*}
Where $|\S|=s$. The size of a mini-file $W_{n,\S}^j$ is $L\left(W_{n,\S}^j\right)\approxeq\left((M/N)^s(1-M/N)^{K-s}f\right)/{K-s-1\choose L-1}$symbols. However the above partition is possible only for $s\in\{s:s<K-L+1\}$.

\subsection{Content Delivery Strategy:}      
For any $s\in\{s:s\leq K-L+1\}$, consider an arbitrary $(s+L-1)$ subset of users denoted by $\B$ (i.e., $\B\subseteq\K,|\B|=s+L-1$). We have ${s+L-1\choose s}$ subsets of $\B$ with length $s$, which we denote as $\S_i$, for $i=1,...,{s+L-1\choose s}$. Now we assign a $L\times1$ vector $\textbf{u}_{\B}^{\S_i}$ to each $\S_i$ such that
\begin{align} \label{bfcon}
&\textbf{u}_{\B}^{\S_i} \perp \textbf{h}_k\text{for all }k\in \B\setminus \S_i \nonumber \\
&\textbf{u}_{\B}^{\S_i} \not\perp \textbf{h}_k\text{for all }k\in \S_i
\end{align}
Since all the subfiles and mini-files are encoded as $\Wt_{n,\S}^j$ $n$ complex numbers per each symbol, any linear combination (non-zero coefficients) with one unknown can easily be decoded. For each $\S_i$ define,
\begin{align}
G(\S_i) = \L_{k\in \S_i}\left(\Wt_{d_k,\S_i\setminus k}^j\right)
\end{align} 
Where $\Wt_{d_k,\S_i\setminus \{k\}}^j$ is a mini-file which is available in the cache of
all users in $\S_i$ except in $k$. $\L_{k\in \S_i}$ represents a random linear combination of the corresponding mini-files for all $k\in \S_i$. The index $j$ is chosen such
that such mini-files have not been observed in the previous $(s+L-1)$-subsets. Now lets define $N(k,\S\setminus \{k\})$ as the index of the next fresh mini-file required by user $k$, which is present in the cache of users $\S\setminus \{k\}$, then
\begin{align}\label{LC}
G(\S_i) = \L_{k\in \S_i}\left(\Wt_{d_k,\S_i\setminus \{k\}}^{N(k,\S_i\setminus \{k\})}\right)
\end{align}
Now construct the sub message to transmit as,
\begin{align}
\textbf{X}(\B) = \frac{1}{\sqrt{s{s+L-1\choose s}}}\sum_{\S\subseteq\B,|\S|=s}\textbf{u}_{\B}^{\S}G(\S)
\end{align}
Since $\textbf{u}_{\B}^{\S}$ is perpendicular to all $k \in \B\setminus \S$, the received signal at any user $k\in\B$ has ${s+L-2\choose s-1}$ non zero messages. To detect them independently we need to construct ${s+L-2\choose s-1}$ sub messages to transmit,
\begin{align}
G_w(\S_i) =& \L^w_{k\in \S_i}\left(\Wt_{d_k,\S_i\setminus k}^{N(k,\S_i\setminus k)}\right)\nonumber\\
\textbf{X}_w(\B) =&\frac{1}{\sqrt{s{s+L-1\choose s}}} \sum_{\S\subseteq\B,|\S|=s}\textbf{u}_{\B}^{\S}G_w(\S)
\end{align} 
Hence the constructed message block to transmit is,
\begin{align}\label{whblc}
\left[\textbf{X}_1(\B),...,\textbf{X}_{{s+L-2\choose s-1}}(\B)\right]
\end{align}
Now update $N(k,\S\setminus k)$ for those mini-files which have appeared in the linear combinations in \eqref{LC}. Repeat the above procedure for all $s+L-1$-subsets of $\K$. This entire process is need to be done for every $s\in\{s:s\leq K-L+1\}$.

For any $s\notin\{s:s\leq K-L+1\}$, we cannot have $(s+L-1)$ subset of users, because it exceeds the maximum number of users $K$. Hence we have no option but to take entire user set as $\B$. Here $\B=\K$. Now we have ${K\choose s}$-subsets of $\K$ of length $s$. Now we have to assign a $L\times1$ vector $\textbf{u}_{\B}^{\S_i}$ to each $\S_i$ such that \eqref{bfcon} follows. For $s\in\{s:s\leq K-L+1\}$ the \eqref{bfcon} has at most one solution. But in this case \eqref{bfcon} has many solutions. We can pick one of the solutions. The following one is one of the examples.
\begin{align}\label{bstu}
\textbf{u}_{\B}^{\S_i} = \arg\max_{\textbf{u}} \min_{k\in \S_i}\textbf{h}_k^H\textbf{u} \hspace{3mm} \text{s.t } \textbf{h}_k^H\textbf{u}=0 \forall k \in \B\setminus \S_i
\end{align}  
Where $\mathbf{H}_{\S}\in\mathbb{C}^{L\times|\S|}$ is the channel matrix $\mathbf{H}$ containing the only columns in $\S$. All the remaining procedure is same with simple substitution $s+L-1=K$, and we have to repeat this procedure for every $s\notin\{s:s\leq K-L+1\}$.

\section{System's symmetric rate  - Linear Combinations in Complex Field} \label{srlccf}
For any user $k\in\K$, and for any $s\in\{s:s\leq K-L+1\}$, there exists at least one $\B\subseteq\K$ such that $k\in\B$. Lets define $\S_k^{\B} = \{\S:\S\subseteq\B,|\S|=s,k\in \S\}$. Then the user $k$ will receive $\textbf{h}_k^H\textbf{X}_w(\B)$
\begin{align}
=& \frac{1}{\sqrt{s{s+L-1\choose s}}}\sum_{\S_k^{\B}}\left(\textbf{h}_k^H\textbf{u}_{\B}^{\S}\sum_{i\in \S}\L_{i\in \S}\left(\Wt_{d_i,\S\setminus \{i\}}\right)\right)\label{rcvd1} \\
=& \frac{1}{\sqrt{s{s+L-1\choose s}}}\sum_{\S_k^{\B}}\left(\textbf{h}_k^H\textbf{u}_{\B}^{\S}\Wt_{d_k,\S\setminus \{k\}}\right)\label{rcvd2}
\end{align}	 
\eqref{rcvd1} is straight forward to verify, and \eqref{rcvd2} follows from the fact that this user has cached and thus can decode the linear combination. Like the above, we will receive for all $w=1,...,{s+L-2\choose s-1}$. Lets say $v={s+L-2\choose s-1}$. The entire received signal block can be represented as, $\textbf{L}_k^{\B}\textbf{W}_k^{\B}+\textbf{z}_k$. Where,
\begin{align}
\textbf{W}_k^{\B}=&\left[\Wt_{d_k,\S_1\setminus k},...,\Wt_{d_k,\S_v\setminus k}\right]^T.\nonumber\\
\textbf{L}_k^{\B}=&\sqrt{\frac{v}{s{s+L-1\choose s}}}U\text{diag}\left(\textbf{h}_k^H\textbf{u}_{\B}^{\S_1},...,\textbf{h}_k^H\textbf{u}_{\B}^{\S_v}\right)\nonumber
\end{align}
$U$ is a $v\times v$ unitary matrix and $\textbf{z}_k$ is additive white Gaussian noise. Now, user $k$ can decode all its required data if the transmission rate is less than $R(\B,k,s)$ which is,
\begin{align}
 \log\left(1+\frac{P_{\max}}{s+L-1}\min_{\S:\S\subseteq\B,|\S|=s,k\in \S}|\textbf{h}_k^H\textbf{u}_{\B}^{\S}|^2\right)
\end{align}
To all the users in $\B$ to successfully decode, they should have the common rate $R(\B,s)$ which is,
\begin{align}\label{CR}
 	\log\left(1+\frac{P_{\max}}{s+L-1}\min_{k\in \B}\min_{\S:\S\subseteq\B,|\S|=s,k\in \S}|\textbf{h}_k^H\textbf{u}_{\B}^{\S}|^2\right)
\end{align}
It is easy to verify that for any $s\notin\{s:s\leq K-L+1\}$, the common rate $R(\B,s)$ is same as in \eqref{CR} with simple substitution $s+L-1=K$.  Since each user in $\B$ decodes approximately $Z(s){s+L-2\choose s-1}F/{K-s\choose L-1}$ bits after transmission to $\B$ is concluded ($Z(s) \approxeq (M/N)^{s-1}(1-M/N)^{K-s+1}$ with high probability \cite{maddah2015decentralized}), the symmetric rate can be written as $R_{sym}\approxeq R^{-1}$, where 
\begin{align*}
R = &\sum_{s=1}^{K-L+1}\frac{Z(s){s+L-2\choose s-1}}{{K-s\choose L-1}}\sum_{\B:\B\subseteq\K,|\B|=s+L-1}\left(R(\B,s)\right)^{-1}\\
&+\sum_{s=K-L+2}^{K}Z(s){K-1\choose s-1}\sum_{\B:\B\subseteq\K,|\B|=s+L-1}\left(R(\B,s)\right)^{-1}				
\end{align*}

\section{System's symmetric rate - Linear Combinations in Finite Field}\label{srlcff}
For any user $k\in\K$, and for any $s\in\{s:s\leq K-L+1\}$, there exists at least one $\B\subseteq\K$ such that $k\in\B$. Lets define $\S_k^{\B} = \{\S:\S\subseteq\B,|\S|=s,k\in \S\}$. Then the user $k$ will receive $\textbf{h}_k^H\textbf{X}_w(\B)$
\begin{align}
=& \frac{1}{\sqrt{{s+L-1\choose s}}}\sum_{\S_k^{\B}}\left(\textbf{h}_k^H\textbf{u}_{\B}^{\S}\sum_{i\in \S}\psi\left(\oplus_{i\in \S}\left(W_{d_i,\S\setminus \{i\}}\right)\right)\right)\label{rcvdff1} \\
=& \frac{1}{\sqrt{{s+L-1\choose s}}}\sum_{\S_k^{\B}}\left(\textbf{h}_k^H\textbf{u}_{\B}^{\S}W_{d_k,\S\setminus \{k\}}\right)\label{rcvdff2}
\end{align}	 
\eqref{rcvdff1} is straight forward to verify, and \eqref{rcvdff2} follows from the fact that this user has cached and thus can decode the linear combination. User $k$ is interested in decoding ${s+L-2\choose s-1}$ subfiles in the above transmission.  Effective sum rate $R_{eff}(\B,k,s)$ for user $k$ in the achievable capacity region of MAC channels is written as,
\begin{align*}
 \min\Biggr[\log\left(1 + \frac{1}{{s+L-1\choose s}}\sum_{\S_k^{\B}}|\textbf{h}_k^H\textbf{u}_{\B}^{\S}|^2P_{\max}\right), \nonumber \\
{s+L-2\choose s-1} \min_{\S_k^{\B}}\log\left(1 + \frac{1}{{s+L-1\choose s}} |\textbf{h}_k^H\textbf{u}_{\B}^{\S}|^2P_{\max}\right)\Biggr].
\end{align*}
And the common rate is written as, $R(\B,s) = \min_{k\in\B}R_{eff}(\B,k,s)$. It is easy to verify that for any  $s\notin\{s:s\leq K-L+1\}$ the common rate is same with $s+L-1=K$. And the symmetric rate can be written as $R_{sym}\approxeq R^{-1}$, where
\begin{align*}
R = &\sum_{s=1}^{K-L+1}\frac{Z(s){s+L-2\choose s-1}}{{K-s\choose L-1}}\sum_{\B:\B\subseteq\K,|\B|=s+L-1}\left(R(\B,s)\right)^{-1}\\
&+\sum_{s=K-L+2}^{K}Z(s){K-1\choose s-1}\sum_{\B:\B\subseteq\K,|\B|=s+L-1}\left(R(\B,s)\right)^{-1}				
\end{align*}
	    		
\footnotesize
\bibliographystyle{IEEEtran}
\bibliography{IEEEabrv,references}

\end{document}